\documentclass[
    ,final            % use final for the camera ready runs
  ]
  {aipproc}

\layoutstyle{8x11single}

\usepackage{bm}
\begin{document}

\title{Heat Flux in a Granular Gas}


\author{J.J. Brey}{
  address={F\'{\i}sica Te\'{o}rica, Universidad de Sevilla, Apartado de Correos 1065, 41080 Sevilla, Spain}
}

\author{M.J. Ruiz-Montero}{
  address={F\'{\i}sica Te\'{o}rica, Universidad de Sevilla, Apartado de Correos 1065, 41080 Sevilla, Spain}
}

\begin{abstract}
A peculiarity of the hydrodynamic Navier-Stokes equations for a granular gas is the modification of the Fourier law, with the presence of an additional contribution to the heat flux that is proportional to the density gradient. Consequently, the constitutive relation involves, in the case of a one-component granular gas, two transport coefficients: the usual (thermal) heat conductivity and a diffusive heat conductivity.  A very simple physical interpretation of this effect, in terms of the mean free path and the mean free time is provided. It leads to the modified Fourier law with an expression for the diffusive Fourier coefficient that differs in a factor of the order of unity from the expression obtained by means of the inelastic Boltzmann equation. Also, some aspects of the Chapman-Enskog computation of the  new transport coefficients as well as of the comparison between simulation results and theory are discussed.

\end{abstract}

\maketitle

\section{Introduction}
\label{s1}
Granular fluids are of increasing interest to several scientific communities. This is partially due to their great relevance in many industrial issues of agricultural, pharmaceutical and chemical significance for packing and transport of grains. Moreover, they raise a number of important conceptual challenges. A class of these questions, widely investigated in the last two decades or so, involve the form and conditions for a hydrodynamic description, similar to the one for normal, molecular fluids \cite{Ha83,Ca90,Go03,DyB06}. In most of the cases, the simplest model of a granular gas at the particle level of description, an ensemble of inelastic hard spheres or disks \cite{BDyS97}, is considered. This model has proven to be able to describe at a qualitative level many of the peculiarities exhibited by real granular gases.

The Navier-Stokes equation for the energy of a granular gas exhibits two main peculiarities when compared with that for normal fluids. First, there is a source term associated to the energy dissipation in collisions. Its origin can be easily understood on the basis of mean field arguments. The other differential feature is the modification of the Fourier law for the heat flux, including a new term that couples it to the density gradient. This new contribution to the heat flux was predicted by kinetic theory methods \cite{BMyD96,BDKyS98,SyG98}, and later on measured in molecular dynamic simulations \cite{SMyR99}. Also, some implications of its existence have been observed in experiments \cite{CyR91,WHyP01,ByK03,HYCMyW04}. The theoretical formal reasons why this density gradient contribution to the heat flux is present in granular fluids and absent in normal ones has been investigated. They are related to the time reversal invariance of both the dynamics of the system and the equilibrium distribution function of normal fluids \cite{Du07}.

On the other hand, although some physical interpretations of the modification of the Fourier law in granular gases have been given, it seems instructive to provide simple intuitive arguments leading to the new term in the Fourier law. In particular, one of the questions addressed here is to show that mean free path and collision time arguments are enough as to predict the presence of the density gradient contribution to the heat flux for granular fluids, in a similar way as they predict all the transport coefficients in molecular gases, aside from a numerical factor.

Another aim of this paper is to discuss a new approximation method to compute Navier-Stokes transport coefficients of dilute gases starting from their Green-Kubo representation. The method is both simple and transparent. It will be illustrated here for the two coefficients appearing in the modified Fourier law. In the elastic limit the results are the same as those derived by the usual Sonine expansion to lowest order, while in the inelastic limit both results are very close.

The remaining of the paper is organized as follows. In the next section,
the derivation of the generalized Fourier law using linear response theory in the context of the Boltzmann equation is shortly reviewed. The relevance of the hydrodynamic part of the spectrum of the linearized Boltzmann operator is emphasized. The formal expressions for the transport coefficients are explicitly evaluated in Sec. \ref{s4} by using an approximation related with the non-hydrodynamic part of the spectrum of the linear Boltzmann operator. The results are shown to be practically equivalent to those that have been obtained by expanding in Sonine polynomials keeping only the lowest order.

In Sec. \ref{s3} a much simpler derivation of the modified Fourier law is carried out. It is based on the concepts of mean free time and mean collision time in a dilute gas, generalizing the arguments used in elementary kinetic theory of molecular gases \cite{Re08}. The origin of both contributions to the heat flux, as well as the way in which the inelasticity modifies the usual  term proportional to the temperature gradient, show up in a  transparent way. Finally, the last section contains a short discussion of some macroscopic effects associated with the new term in the expression of the heat flux. The specific state considered corresponds to a series of experiements carried out by different groups.

\section{Linearized Boltzmann equation and hydrodynamics}
\label{s2}
The system considered is a dilute granular gas of smooth inelastic hard spheres ($d=2$) or disks ($d=3$) of mass $m$ and diameter $\sigma$. The coefficient of normal restitution $\alpha$ will be taken as a velocity-independent constant, although the theory has also been extended to more realistic models in which it depends on the relative velocity of the colliding particles \cite{ByP03}. The one-particle distribution function of the system $f({\bm r},{\bm v},t)$  obeys the inelastic nonlinear Boltzmann equation \cite{LSJyCh84,GyS95}. This equation has a particular solution describing the homogeneous cooling state (HCS),
\begin{equation}
\label{2.2}
f_{H}({\bm v},t)=n v_{0}^{-d}(t) \chi (c), \quad \quad v_{0} \equiv \left( \frac{2T_{H}}{m} \right)^{1/2},
\quad \quad {\bm c} \equiv \frac{{\bm v}}{v_{0}(t)},
\end{equation}
where $n$ is the uniform number density,  $\chi$ is an isotropic function of the scaled velocity ${\bm c}$, and $T_{H}$  is the uniform  temperature of the system. It decays monotonically in time according with the Haff law,
\begin{equation}
\label{2.3}
\frac{\partial T_{H}}{\partial t}  =- \zeta_{H} T_{H}(t).
\end{equation}
Approximated expressions for the distribution $\chi (c)$ and for the cooling rate $\zeta_{H}$ have been obtained by expanding the function in Sonine polynomials and keeping only the lowest order polynomial \cite{GyS95,vNyE98}. The approximated distribution function has the form
\begin{equation}
\label{2.4}
\chi(c)= \frac{e^{-c^{2}}}{\pi^{d/2}}\, \left[ 1
+a_{2}(\alpha) S^{(2)} (c^{2}) \right],
\end{equation}
where
\begin{equation}
\label{2.5}
S^{(2)}(c^{2})= \frac{c^{4}}{2}-\frac{d+2}{2}\, c^{2} +\frac{d(d+2)}{8}
\end{equation}
and
\begin{equation}
\label{2.6}
a_{2}(\alpha)= \frac{16(1-\alpha)(1-2 \alpha^{2})}{9+24d+(8d-41)\alpha+30 \alpha^{2}
-30 \alpha^{3}}\, .
\end{equation}
The expression for the cooling rate is:
\begin{equation}
\label{2.7}
\zeta_{0} \equiv \frac{\zeta_{H} \ell}{v_{o}(t)}= \frac{ \sqrt{2} \pi^{(d-1)/2} (1-\alpha^{2})}{\Gamma \left(d/2 \right) d} \left[ 1+ \frac{3 a_{2}(\alpha)}{16} \right].
\end{equation}
Here $\ell \equiv (n \sigma^{d-1})^{-1}$ is proportional to the mean free path of the gas. Suppose now a small perturbation around the HCS, and define $\delta f$ by
\begin{equation}
\label{2.7a}
f({\bm r},{\bm v},t)= f_{H}({\bm v},t)+ \delta f ({\bm r},{\bm v},t), \quad \quad |\delta f({\bm r},{\bm v},t)\ \ll |f_{H}({\bm v},t)|.
\end{equation}
To eliminate the time dependence associated to the HCS, it is convenient to introduce dimensionless length $l$ and time scales $s$ defined by
\begin{equation}
\label{2.8}
{\bm l} \equiv \frac{\bm r}{\ell}, \quad \quad s \equiv \int_{0}^{t}dt_{1} \frac{v_{0}(t_{1})}{\ell},
\end{equation}
respectively. The time scale $s$ is proportional to the accumulated average number of collisions per particle. The dimensionless form of the deviation of the distribution function is
\begin{equation}
\label{2.9}
\delta \chi ({\bm l},{\bm c},s) \equiv n^{-1} v_{0}^{d}(t) \delta f ({\bm r},{\bm v},t).
\end{equation}
This function obeys the linear equation \cite{BDyR03,ByD05}
\begin{equation}
\label{2.10}
\left(\frac{\partial}{\partial s}+ {\bm c} \cdot \frac{\partial}{\partial {\bm l}} \right) \delta \chi ({\bm l},{\bm c},s)= \Lambda ({\bm c}) \chi ({\bm l},{\bm c},s).
\end{equation}
The linear Boltzmann operator $\Lambda$ is given by
\begin{equation}
\label{2.11}
\Lambda({\bm c}_{1}) \equiv \int d {\bm c}_{2}\,
\overline{T}_{0}({\bm c}_{1},{\bm c}_{2}) (1+P_{12}) \chi
({c}_{2})-\frac{\zeta_{0}}{2} \frac{\partial}{\partial {\bm c}_{1}}
\cdot {\bm c}_{1}.
\end{equation}
where $\overline{T}_{0}({\bm c}_{1},{\bm c}_{2})$ is the dimensionless binary collision operator for inelastic hard spheres or disks,
\begin{equation}
\label{2.12}
\overline{T}_{0}({\bm c}_{1},{\bm c}_{2})=  \int d
\widehat{\bm \sigma}\, \Theta ({\bm c}_{12} \cdot
\widehat{\bm \sigma}) {\bm c}_{12} \cdot \widehat{\bm
\sigma}   \left[ \alpha^{-2} b_{\bm \sigma}^{-1}({\bm c}_{1},{\bm
c}_{2}) -1 \right].
\end{equation}
Here ${\bm c}_{12} \equiv {\bm c}_{1}- {\bm c}_{2}$,\,  $d
\widehat{\bm \sigma}$ is the solid angle element for the unit vector $\widehat{\bm \sigma}$,
$\Theta$ is the Heaviside step function, and $  b_{\bm \sigma}^{-1}({\bm c}_{1},{\bm
c}_{2})$ is an operator changing all the functions of ${\bm c}_{1}$ and ${\bm c}_{2}$ to its right by the
same functions of the precollisional velocities ${\bm c}^{*}_{1}$ and ${\bm c}^{*}_{2}$, given by
\begin{eqnarray}
\label{2.13} {\bm c}^{*}_{1} \equiv b_{\bm \sigma}^{-1} {\bm c}_{1}=
{\bm c}_{1}-\frac{1+\alpha}{2 \alpha} ( \widehat{\bm \sigma} \cdot
{\bm c}_{12} ) \widehat{\bm \sigma},
\nonumber \\
{\bm c}^{*}_{2} \equiv b_{\bm \sigma}^{-1} {\bm c}_{2}= {\bm
c}_{2}+\frac{1+\alpha}{2 \alpha} ( \widehat{\bm \sigma} \cdot {\bm
c}_{12} ) \widehat{\bm \sigma}.
\end{eqnarray}
Solutions to Eq.\ (\ref{2.10}) are sought in a Hilbert space defined by the scalar product
\begin{equation}
\label{2.14}
\langle g|h\rangle \equiv  \int d{\bm c}\, \chi^{-1}(c) g^{*}({\bm c}) h({\bm c}),
\end{equation}
with $g^{*}({\bm c})$ being the complex conjugate of $g({\bm c})$. It is then useful to consider the homogeneous eigenvalue problem
\begin{equation}
\label{2.15}
\Lambda ({\bm c}) \xi_{\beta} ({\bm c})= \lambda_{\beta} \xi_{\beta} ({\bm c}).
\end{equation}
The solutions of this equation corresponding to the infinite wavelength limit of the hydrodynamic
equations are given by \cite{BDyR03,ByD05}
\begin{equation}
\label{2.16}
\lambda_{1}=0, \quad \lambda_{2}=\frac{\zeta_{0}}{2}\, ,
\quad \lambda_{3} =-\frac{\zeta_{0}}{2},
\end{equation}
\begin{equation}
\label{2.17} \xi_{1}({\bm c})= \chi(c)+\frac{\partial}{\partial {\bm
c}} \cdot \left[ {\bm c} \chi (c) \right], \quad {\bm \xi}_{2}({\bm
c})=-\frac{\partial \chi(c)}{\partial {\bm c}}, \quad \xi_{3}({\bm
c})= -\frac{\partial}{\partial {\bm c}} \cdot \left[ {\bm c} \chi
(c) \right].
\end{equation}
The eigenvalue $\lambda_{2}$ is $d$-fold degenerated. The operator $\Lambda ({\bm c})$ is not Hermitian and the eigenfunctions $\xi_{i}$ are not orthogonal. This leads to introduce a set of functions $\overline{\xi}_{i}$ that be orthogonal to the above eigenfunctions. A convenient choice is
\begin{equation}
\label{2.18}
\label{3.21a} \overline{\xi}_{1}({\bm c})=\chi(c), \quad \overline{\bm
\xi}_{2} ({\bm c})={\bm c} \chi(c), \quad \overline{\xi}_{3} ({\bm
c})= \left( \frac{c^{2}}{d} +\frac{1}{2} \right) \chi(c); \quad \quad
\langle \overline{\xi}_{\beta} |\xi_{\beta^{\prime}}
\rangle =\delta_{\beta,\beta^{\prime}}\, .
\end{equation}
The general solution of the linearized Boltzmann equation in the Hilbert space can be formally written in the Fourier representation as
\begin{equation}
\label{2.19}
\delta \chi =\sum_{\beta} a_{\beta}({\bm k},s)  \xi_{\beta}({\bm k},{\bm c}) + \delta \chi^{m} ({\bm k},{\bm c},s),
\end{equation}
where the sum extends over the $d+2$ hydrodynamic modes of the operator $\Lambda-i {\bm k} \cdot {\bm c}$ and $\delta \chi^{m}$ contains all the other ``microscopic'' modes. The hydrodynamic modes for finite ${\bm k}$ of $\Lambda-i {\bm k} \cdot {\bm c}$ are defined as those eigenvalues that are continuously connected as functions as ${\bm k}$ to those given in Eqs.\ (\ref{2.16}).

It is now {\em assumed} that the hydrodynamic part of the spectrum dominates for long times and small gradients. Then, the microscopic part can be neglected in Eq.\ (\ref{2.19}). Then, the formal solution of Eq.\ (\ref{2.10}) to first order in $k$ can be expressed as \cite{BDyR03}
\begin{equation}
\label{2.20}
\delta \chi ({\bm k},{\bm c},s) = e^{s\Lambda} \delta \chi ({\bm k},{\bm c},0) - \sum_{\beta} \left[ \int_{0}^{s} ds^{\prime}\, e^{s^{\prime} \left( \Lambda -\lambda_{\beta} \right)} {\bm c} \xi_{\beta}({\bm c}) \right] \cdot i {\bm k} a_{\beta}({\bm k},s).
\end{equation}
Moreover, the coefficients $a_{\beta}$ in the above expression can be identified as
\begin{equation}
\label{2.20a}
a_{\beta}({\bm k},s) \simeq \langle \overline{\xi}_{\beta}|\delta \chi ({\bm k},s) \rangle.
\end{equation}
These coefficients can be expressed as functions of the hydrodynamic fields density, velocity, and temperature. In this way, a formal expression for $\delta  \chi$ valid to first order in the gradients ($k$) is obtained and, using it, the Navier-Stokes expressions for the heat flux and the pressure tensor. They are a generalization of the well-known Green-Kubo formulas for molecular gases. The details of the calculations are given in ref. \cite{BDyR03}. In the next section the expression for the heat flux flux will analyzed in some detail.

\section{Green-Kubo expression for the heat flux}
\label{s4}
\begin{center}
\begin{figure}
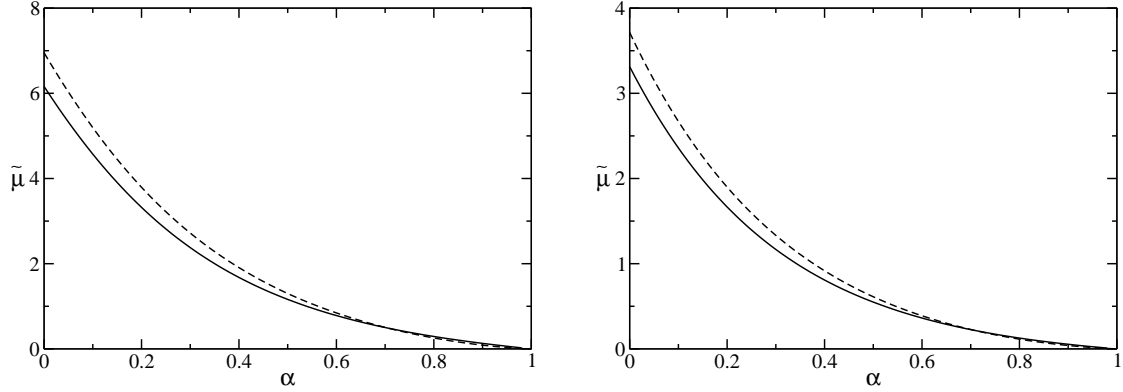

\includegraphics[scale=0.3]{byr12af1a.eps} \quad \quad
\includegraphics[scale=0.3]{byr12af1b.eps}
  \caption{Diffusive heat conductivity of a dilute granular gas of hard disks (left side) and spheres (right side) as a function of the coefficient of normal restitution $\alpha$. The solid lines are the results found in the first Sonine approximation, while the dashed lines have been obtained using Eq.\ (\protect{\ref{4.13}}).}
\label{fig1}
\end{figure}
\end{center}

Following the procedure sketched in the previous section, it is obtained that the heat flux ${\bm q}({\bm r},t)$ to Navier-Stokes order has the form
\begin{equation}
\label{4.1}
{\bm q} = -\kappa {\bm \nabla T} - \mu {\bm \nabla} n,
\end{equation}
where $\kappa$ is the (thermal) heat conductivity and $\mu$ a transport coefficient vanishing in the elastic limit and that is referred to as the diffusive heat conductivity. Their formal expressions are
\begin{equation}
\label{4.2}
\kappa = n m \ell v_{0}(t) \widetilde{\kappa}(s), \quad \quad \mu= m \ell v_{0}^{3}(t) \widetilde{\mu}(s),
\end{equation}
\begin{equation}
\label{4.3}
\widetilde{\kappa}(s) = \frac{1}{d} \int d{\bm c}\,  {\bm \Sigma} ({\bm c}) \cdot {\bm \Phi}_{3} ({\bm c},s),
\end{equation}
\begin{equation}
\label{4.4}
\widetilde{\mu}(s) = \frac{1}{d} \int d{\bm c}\,  {\bm \Sigma} ({\bm c}) \cdot {\bm \Phi}_{1} ({\bm c},s),
\end{equation}
with
\begin{equation}
\label{4.5}
{\bm \Sigma}({\bm c}) \equiv \left( c^{2}-
\frac{d+2}{2} \right) {\bm c}.
\end{equation}
\begin{equation}
\label{4.6}
{\bm \Phi}_{1} ({\bm c},s) = \int_{0}^{s} ds^{\prime}\, e^{s^{\prime} \Lambda ({\bm c})} \xi_{1} ({\bm c}) {\bm c}
+2 {\bm \Phi}_{3} ({\bm c},s),
\end{equation}
\begin{equation}
\label{4.7}
{\bm \Phi}_{3} ({\bm c},s)= \frac{1}{2} \int_{0}^{s} ds^{\prime}\, e^{s^{\prime} \left[ \Lambda ({\bm c}) + \zeta_{0}/2 \right]} {\xi}_{3} ({\bm c}) {\bm c}.
\end{equation}
Here $\xi_{1}({\bm c})$ and $\xi_{2} ({\bm c})$ are the eigenfunctions given by Eqs.\ (\ref{2.17}). As usual, $\widetilde{\kappa}(s)$ and $\widetilde{\mu}(s)$ are expected to reach steady plateau values for large enough $s$, when the hydrodynamic description is accurate. Both transport coefficients have been evaluated in the first Sonine approximation \cite{BDKyS98,ByC01}. Here an alternative approximation will be discussed \cite{BMyG11}. It consists in  treating ${\bm \Sigma (c)} \chi (c)$ as an eigenfunction of the operator adjoint of $\Lambda$, $\Lambda^{+}$, i.e. it is considered that
\begin{equation}
\label{4.8}
\Lambda^{+} ({\bm c}) {\bm \Sigma} ({\bm c}) \chi (c) \simeq \overline{\lambda}_{5} {\bm \Sigma} ({\bm c}) \chi (c),
\end{equation}
and $\overline{\lambda}_{5}$ is obtained by multiplying this equation by $c_{x}$ and afterwards integrating over the velocity ${\bm c}$. The results reads
\begin{equation}
\label{4.9}
\overline{\lambda}_{5} = \frac{ 4 J(\alpha)}{(d+2) a_{2} (\alpha)} + \frac{\zeta_{0}(\alpha)}{a_{2}(\alpha)} + \frac{3 \zeta_{0} (\alpha)}{2}\,  ,
\end{equation}
\begin{equation}
\label{4.10}
J(\alpha)  =  - \frac{\pi^{(d-1)/2}(1+\alpha)}{32 \sqrt{2} d \Gamma \left(d/2 \right)}
\left\{16(2+d) (1-\alpha)+a_{2}(\alpha) \left[ 70+47d -3(34+5d) \alpha \right] \right\}.
\end{equation}
Note that assuming Eq.\ (\ref{4.8}) leads to an exponential decay of the time correlation function involved in the calculation of $\widetilde{\kappa}$ (see Eqs.\ (\ref{4.3}) and (\ref{4.7})). With this approximation, the long time limit value of the thermal heat conductivity $\widetilde{\kappa}(s)$ is given by
\begin{equation}
\label{4.11}
\widetilde{\kappa} \simeq \frac{1}{2d} \int_{0}^{\infty} ds^{\prime} \int d{\bm c}\, e^{s^{\prime} \left( \overline{\lambda}_{5} + \zeta_{0}/2 \right)} \xi_{3}({\bm c}) {\bm \Sigma} ({\bm c}) \cdot {\bm c}
= - \frac{1}{2d} \left( \overline{\lambda}_{5} + \frac{\zeta_{0}}{2} \right)^{-1}
\int d{\bm c}\, \xi_{3}({\bm c}) {\bm \Sigma} ({\bm c}) \cdot {\bm c}.
\end{equation}
Evaluation of the velocity integral using Eqs.\ (\ref{2.17}) and (\ref{2.4}) yields
\begin{equation}
\label{4.12}
\widetilde{\kappa} \simeq \frac {(d+2) \left[ 1+ 2 a_{2} (\alpha) \right]}{ 2 \left( 2 |\overline{\lambda}_{5}|-\zeta_{0} \right)}.
\end{equation}
By proceeding in an analogous way, the expression for the diffusive heat conductivity can be  evaluated giving
\begin{equation}
\label{4.13}
\widetilde{\mu} \simeq 2 \widetilde{\kappa} - \frac{ (d+2) \left[ 2+a_{2} (\alpha) \right] }{4 | \overline{\lambda}_{5}|}\, .
\end{equation}
As mentioned above, these coefficients have also been computed in the first Sonine approximation \cite{BDKyS98,ByC01}. The results obtained in that way are almost indistinguishable from Eqs. (\ref{4.12}) and (\ref{4.13}), as it can be seen in Fig. \ref{fig1}  for the coefficient of diffusive heat conductivity
$\mu$. Similar results are obtained for the coefficient of thermal heat conductivity $\kappa$. It is worth to mention that a modified Sonine expansion, in which the Gaussian is replaced by the (approximated) distribution of the HCS, has also been employed to compute the transport coefficients of a dilute granular gas \cite{GSyM07}. The results reported are not equivalent to Eqs.\ (\ref{4.12}) and (\ref{4.13}), although they are very close for $\alpha \geq 0.65$.

\section{Elementary derivation of the modified Fourier law}
\label{s3}
Suppose a three-dimensional granular gas in an arbitrary time-dependent state, but with macroscopic gradients only in the $z$-direction. It has already been mentioned  that there is a homogeneous reference state, the homogeneous cooling state, in which the temperature of the system decays monotonically in time according to the Haff law. The interest here is to derive an expression for the heat flux occurring in the system in the direction of the gradients when  they are small. Since the heat flux is by definition the flux of internal energy that is not associated to a macroscopic flux of mass, it will be assumed that the average flux of particles vanishes everywhere in the system.

\begin{figure}
  \includegraphics[scale=0.5]{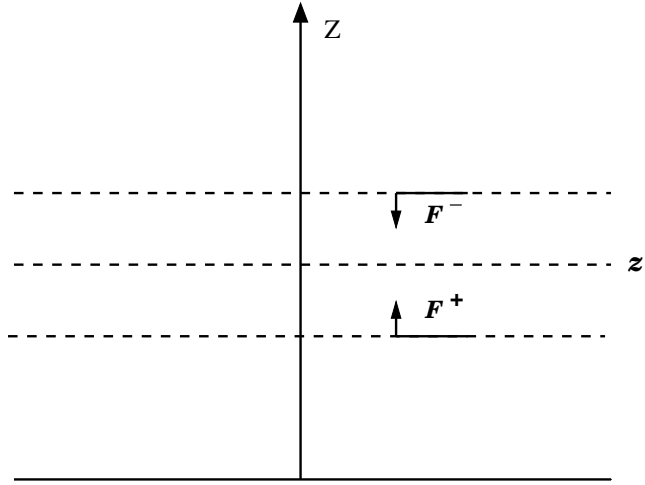}
  \caption{Sketch of the way in which the fluxes through a plane $z=$ constant in the interior of the granular gas are computed.}
  \label{fig2}
\end{figure}

Consider a plane $z=$ constant inside the gas (see Fig. \ref{fig2}). In the dilute limit, the flux of particles crossing it in the direction of increasing $z$ can be estimated as
\begin{equation}
\label{3.1}
{\cal F}^{(+)}(z,t) \approx \frac{1}{4}\, n^{(-)} (z,t) \overline{v}^{(-)} (z,t).
\end{equation}
Here $n^{(-)}(z,t)$ and $\overline{v}^{(-)}(z,t)$ are the number density and average velocity, respectively, of the granular gas, just below the considered plane. Upon writing the above expression, it has been assumed that the velocity distribution in that region can be accurately approximated by an isotropic function. Similarly, the flux of particles through the same plane but in the decreasing $z$-direction is
\begin{equation}
\label{3.2}
{\cal F}^{(-)}(z,t) \approx \frac{1}{4}\, n^{(+)} (z,t) \overline{v}^{(+)} (z,t).
\end{equation}
where now $n^{(+)}(z,t)$ and $\overline{v}^{(+)}(z.t)$ refer to values just above the plane $z=$ constant.

The requirement that there is no net flux of particles implies that
\begin{equation}
\label{3.3}
{\mathcal F}^{(+)}(z,t)= {\mathcal F}^{(-)}(z,t) = {\mathcal F}(z,t) \approx \frac{1}{4} n(z,t) \overline{v}(z,t).
\end{equation}
Next, the average value, $e^{(+)}(z,t)$ of the energy being carried out by a particle crossing the $z$-plane at time $t$ in the positive direction will be estimated. Typically, the last collision suffered by the particle before crossing the plane was at a distance of the order of the local mean free path of the gas, $\lambda (z,t)$. Moreover, that collision took place at a time $t-\tau$, where $\tau$ is the mean free time, i.e. the average time between consecutive collisions of a given particle. In the spirit of mean free path theories, it is assumed that the effect of the collision can be modelled as the thermalization of the velocity distribution of the particles to the local temperature. Consequently, the average energy of a particle crossing the plane in the direction of increasing $z$ is approximated by
\begin{equation}
\label{3.4}
e^{(+)}(z,t) \approx \frac{3}{2} T(z-\lambda,t-\tau).
\end{equation}
In the same way, the average energy carried out by a particle crossing the plane in the direction of decreasing $z$ is estimated as
\begin{equation}
\label{3.5}
e^{(-)}(z,t) \approx \frac{3}{2} T(z+\lambda,t-\tau).
\end{equation}
Consequently, the neat heat flux through the $z$-plane is given by
\begin{equation}
\label{3.6}
q_{z}(z,t) \approx \frac{3}{8} n(z,t) \overline{v}(z,t) \left[ T(z-\lambda, t- \tau)-T(z+\lambda, t - \tau) \right].
\end{equation}
The aim now is to analyze the above expression in the limit of small gradients of the hydrodynamic fields, namely the number density and the temperature. This requires some care when approximating the temperatures difference appearing on the right hand side of the equation. It is convenient to expand the temperature difference as
\begin{equation}
\label{3.8}
 T(z-\lambda, t- \tau)-T(z+\lambda, t - \tau) \approx T(z-\lambda,t)-T(z+\lambda,t)- \tau \frac{\partial}{\partial t} \left[ T(z-\lambda,t)-T(z+\lambda,t) \right] .
\end{equation}
This approximation requires that the temperature profile of the system changes very little over distances of the order of the mean free path $\lambda$ for times of the order of the collision time $\tau$. Given that the difference $T(z-\lambda,t)-T(z+\lambda,t)$  is at least of first order in the gradients, its time derivative must be computed without introducing any additional gradient operator. Therefore,
\begin{equation}
\label{3.9}
\frac{\partial}{\partial t}\, T(z\mp \lambda,T)
 \approx -T(z \mp \lambda,t) \zeta (z\mp \lambda,t),
\end{equation}
where $\zeta(z,t)$ is the cooling rate to zeroth order in the gradients, i.e. the cooling rate of the homogeneous cooling state, but particularized for the fields $n(z,t)$ and $T(z,t)$. In this way, it follows that
\begin{equation}
\label{3.10}
T(z-\lambda, t- \tau)-T(z+\lambda, t - \tau) \approx T(z-\lambda,t)-T(z+\lambda,t)+\tau \left[ T(z-\lambda,t) \zeta(z-\lambda,t)-T(z+\lambda,t) \zeta(z+\lambda,t) \right].
\end{equation}
Next, expansions around $z$, $t$ are carried out as
\begin{equation}
\label{3.11}
T(z-\lambda,t)-T(z+\lambda,t) \approx -2 \lambda \frac{\partial T(z,t)}{\partial z},
\end{equation}
\begin{equation}
\label{3.12}
T(z-\lambda,t) \zeta(z-\lambda,t)-T(z+\lambda,t) \zeta(z+\lambda,t) \approx - 2 \lambda \left[ \zeta (z,t) \frac{\partial T(z,t)}{\partial z} + T(z,t) \frac{\partial \zeta (z,t)}{\partial z} \right].
\end{equation}
Since $\zeta \propto nT^{1/2}$ (see Eq.\ (\ref{2.7})), it is
\begin{equation}
\label{3.13}
\frac{\partial \zeta}{\partial z} = \frac{\zeta}{n} \frac{\partial n}{\partial z}+ \frac{\zeta}{2T} \frac{\partial T}{\partial z}
\end{equation}
and substitution in Eq.\ (\ref{3.12}) yields
\begin{equation}
\label{3.14}
T(z-\lambda,t) \zeta(z-\lambda,t)-T(z+\lambda,t) \zeta(z+\lambda,t) \approx - 2 \lambda \left[ \frac{3 \zeta(z,t)}{2} \frac{\partial T (z,t)}{\partial z}+ \frac{T(z,t) \zeta (z,t)}{n} \frac{\partial n(z,t)}{\partial z} \right].
\end{equation}
Use of Eqs.\ (\ref{3.11}) and (\ref{3.14}) into Eq.\ (\ref{3.10}) leads to
\begin{equation}
\label{3.15}
T(z-\lambda, t- \tau)-T(z+\lambda, t - \tau) \approx
-2 \lambda \left[ 1+\frac{3 \tau}{2}\, \zeta(z,t) \right] \frac{\partial T(z,t)}{\partial z} -
\frac{2 \lambda \tau T(z,t) \zeta (z,t)}{n(z,t)} \frac{\partial n(z,t)}{\partial z}.
\end{equation}
Finally, when the above expression is introduced into Eq.\ (\ref{3.6}), it gives the generalization of the Fourier law,
\begin{equation}
\label{3.15a}
q_{z} = - \kappa \frac{\partial T}{\partial z}- \mu \frac{\partial n}{\partial z}\, ,
\end{equation}
with the following expressions for the thermal heat conductivity and the diffusive heat conductivity transport coefficients:
\begin{equation}
\label{3.16}
\kappa = \frac{3 n \overline{v} \lambda}{4}\, \left( 1+\frac{3 \tau \zeta}{2} \right),
\end{equation}
\begin{equation}
\label{3.17}
\mu = \frac{3}{4}\,  \overline{v} \lambda \tau T \zeta.
\end{equation}
Define
\begin{equation}
\label{3.18}
\kappa_{0} \equiv \frac{3 n \overline{v} \lambda}{4}.
\end{equation}
This is the elastic limit of Eq. (\ref{3.16}) if the (small) dependence of $\overline{v}$ on the inelasticity is neglected. Then, Eqs.\ (\ref{3.16}) and (\ref{3.17}) become
\begin{equation}
\label{3.19}
\kappa(T)= \kappa_{0}(T) \left( 1+\frac{3 \tau \zeta}{2} \right),
\end {equation}
\begin{equation}
\label{3.20}
\mu (T) = \mu_{0}(T)  \tau \zeta,
\end{equation}
where
\begin{equation}
\label{3.21}
\mu_{0}(T) \equiv \frac{T \kappa_{0} (T)}{n}
\end{equation}
has been introduced.

Therefore, the very elementary reasoning presented above leads to and expression for the heat flow in a dilute granular gas having the same structure as the one derived by using more exact and complex methods based on the Boltzmann equation or linear response theories discussed in the previous section. To carry out a more detailed comparison with the explicit forms obtained by the latter methods, define a dimensionless cooling rate by
\begin{equation}
\label{3.22}
\zeta^{*} \equiv \frac{\tau \zeta}{2}.
\end{equation}
In terms of it, Eqs.\ (\ref{3.19}) and (\ref{3.20}) read
\begin{equation}
\label{3.23}
\kappa (T)= \kappa_{0}(T) \kappa^{*}, \quad \quad
\mu (T) = \mu_{0}(T) \mu^{*},
\end{equation}
with
\begin{equation}
\label{3.24}
\kappa^{*} = 1+ 3 \zeta^{*}, \quad \quad   \mu^{*} = 2 \zeta^{*}.
\end{equation}
It is well known that mean free path reasonings leads to values of the elastic heat conductivity $\kappa_{0}$ that differs from those obtained by applying the Chapmann-Enskog procedure to the Boltzmann equation by a factor of the order of unity. This comparison will be not repeated here and attention will be restricted to the values of the reduced coefficients $\kappa^{*}$, and $\mu^{*}$.

The reduced cooling rate $\zeta^{*}$ has been defined in Eq.\ (\ref{3.22}). Note that this is a property of the HCS and, therefore, is assumed to be known. Agreement between the definition of $\zeta_{0}$ in Eq.\ (\ref{2.7}) and $\zeta^{*}$ is obtained if $\tau$ is chosen to be $\tau= 2 \ell/v_{0}(t)$. Moreover, keeping only up to order $1-\alpha$, Eqs. (\ref{4.12}) and (\ref{4.13}) yield
\begin{equation}
\label{3.25}
\kappa^{*} \sim 1+\frac{2d}{d-1} \zeta_{0}(\alpha), \quad \quad \mu^{*} \sim \frac{d}{d-1} \zeta_{0} (\alpha).
\end{equation}
The agreement can be considered as satisfactory, taking into account the looseness of the arguments used.

\section{Heat flux in an open fluidized granular gas}
\label{s5}

\begin{figure}
\includegraphics[scale=0.4]{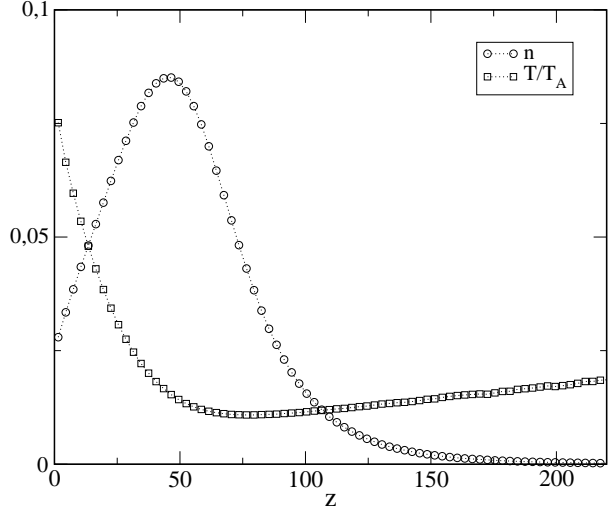}
  \caption{Temperature and density profiles for a two-dimensional system with $\alpha=0.9$ and $N_{z} \equiv N \sigma /W =7$. The temperature is scaled with some arbitrary temperature $T(A)$ and the density is measured in unit of $\sigma^{2}$.}
  \label{fig3}
\end{figure}

When an open granular in presence of gravity is fluidized by means of a vibrating bottom plate, it reaches a steady state in which the energy supplied by the vibrating wall balances the energy dissipated in collisions. Assuming that there are gradients only in the $z$-direction, the Navier-Stokes equations of the systems are \cite{BRyM01}:
\begin{equation}
\label{5.1}
\frac{\partial p}{\partial z} =-nmg,
\end{equation}
\begin{equation}
\label{5.1b}
\frac{2}{nd}\, \frac{\partial q_{z}}{\partial z} + T \zeta=0,
\end{equation}
where $p=nT$ is the hydrodynamic pressure, $\zeta$ is the local cooling rate of the gas, and $g$ is the intensity of the gravitational field in the direction of decreasing $z$. In ref. \cite{BRyM01} it was shown that the temperature profile soltuion of the above equation with the corresponding boundary conditions has a minimum $T_{m}$ at a certain height $z_{m}$, increasing from there on. Therefore, beyond the temperature inversion the heat current flows from cold to hot, in a counterintuitive way. On the other hand, the density profile  exhibits a maximum (at a different position). Using Eq. (\ref{5.1}), the heat flux given in Eq.\ (\ref{4.1}) can be rewritten as
\begin{equation}
\label{5.2}
q_{z}(z)= n \ell v_{0}(t) \left[- \left( \widetilde{\kappa}-2\widetilde{\mu} \right)\frac{\partial T}{\partial z}  + 2 \widetilde{\mu} mg \right].
\end{equation}
At the temperature minimum, this equation becomes
\begin{equation}
\label{5.3}
q_{z}(z_{m}) = 2nm g \ell v_{0}(t) \widetilde{\mu}.
\end{equation}
Note that the existence of the minimum and, therefore, of the increase of the temperature with the height, is directly related with existence of the new term in the Fourier law. Eq.\ (\ref{5.3}) has been used in numerical simulations to measure the exponent $\mu$ and check the theoretical prediction derived from the Boltzmann equation \cite{ByR04}. The Direct Simulation Monte Carlo Methods was used in order to explote the symmetry of the system, rendering the numerical methods much more efficient \cite{Bi94}.  Both two-dimensional and three dimensional gases where studied. As an example, the steady temperature and density profiles for a system of hard disks ($d=2$), with $\alpha=0.9$, and $N \sigma /W= 7$ , where $N$ is the number of particles and $W$ is the width of the system,  are shown in Fig. \ref{fig3}. Other details can be found in ref. \cite{ByR04}. The  lines are guides for the eye. Both the minimum in the temperature profile and the maximum in the density one are clearly identified.

\begin{center}
\begin{figure}
\includegraphics[angle=-90,scale=0.4]{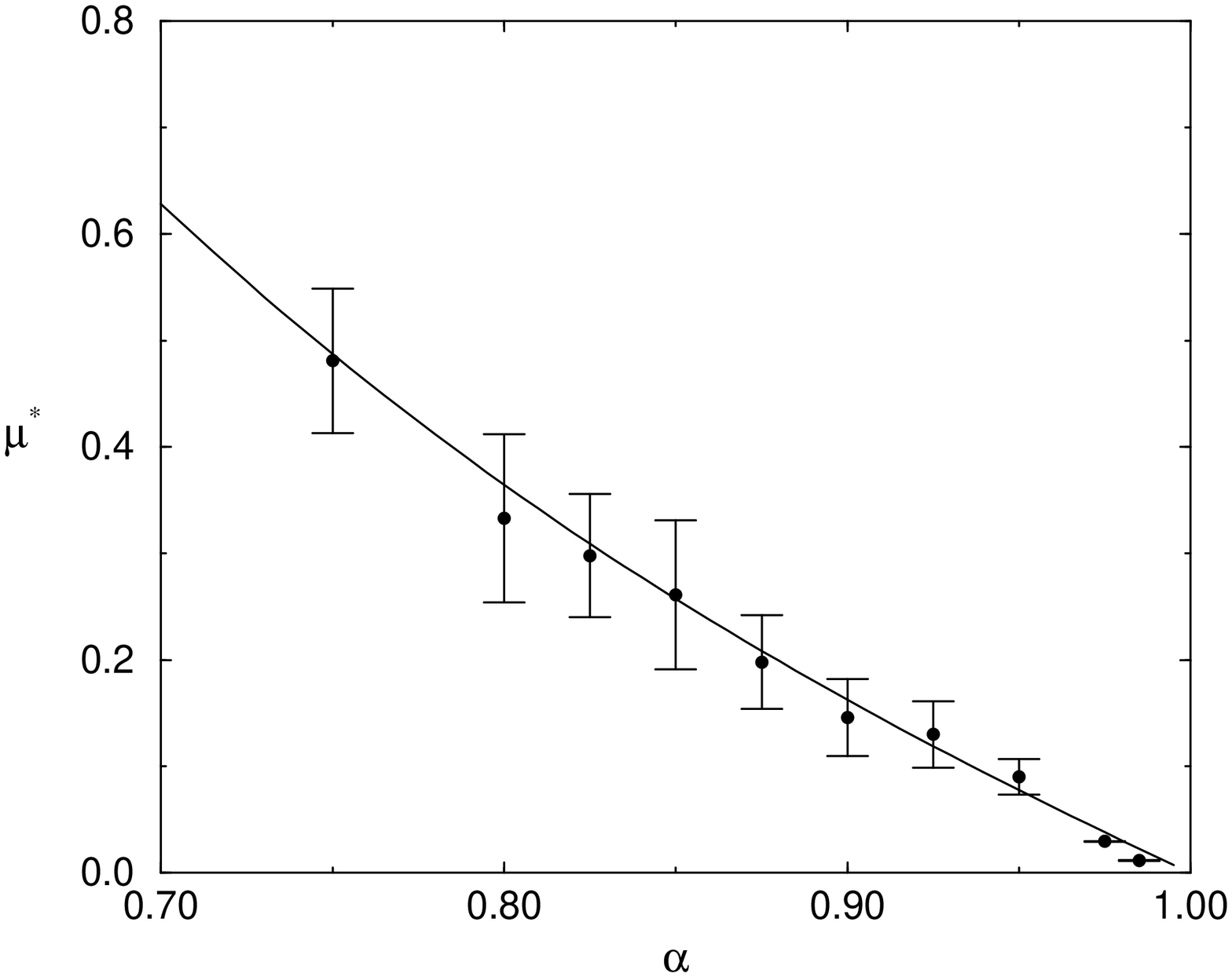} \quad \quad
\includegraphics[angle=-90,scale=0.4]{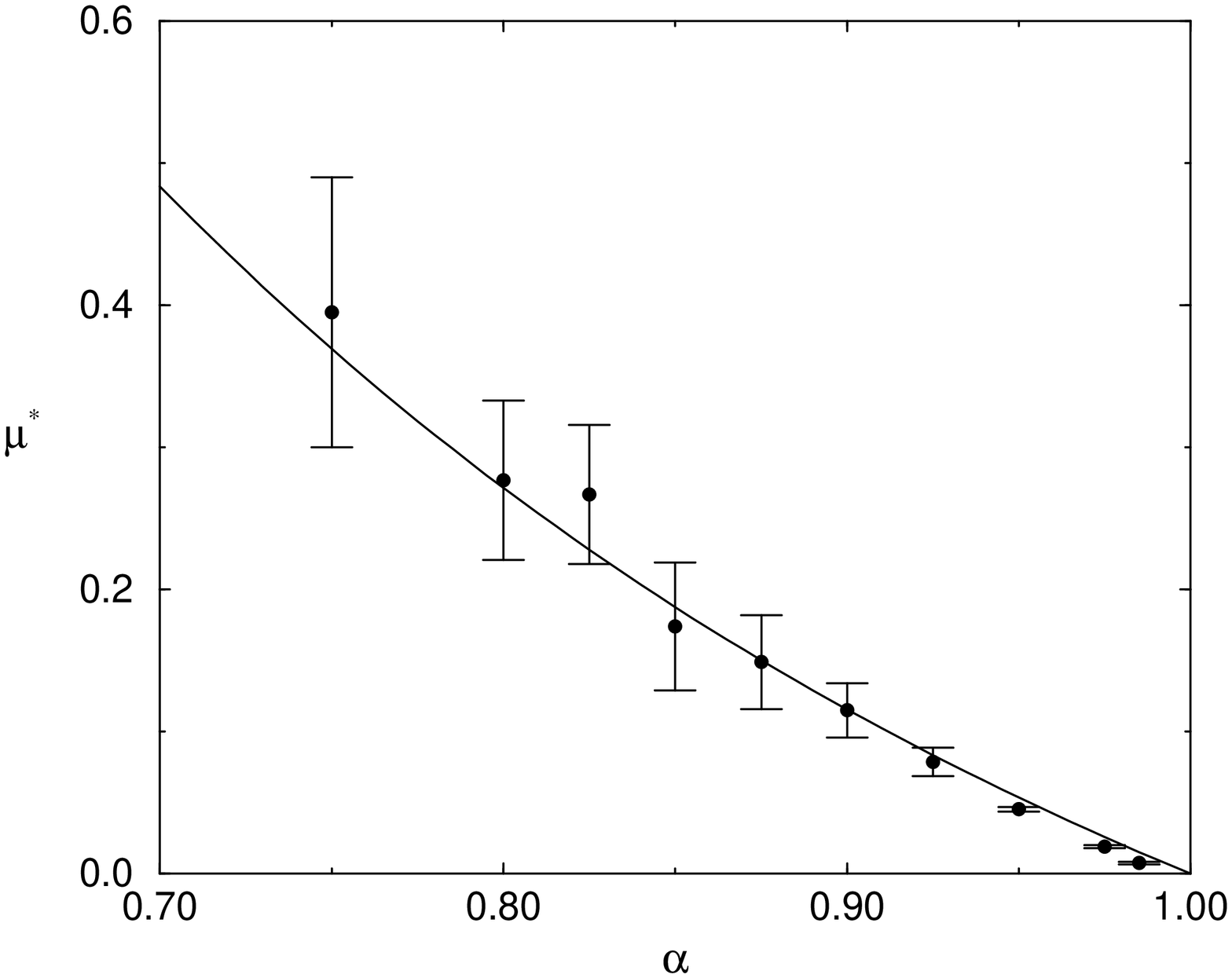}
  \caption{Reduced diffusive heat conductivity $\mu^{*}$ for a dilute gas of inelastic hard disks (left side) and spheres (right side) as a function of the coefficient of normal restitution $\alpha$. the symbols are simulation results and the solid lines the theoretical predictions discussed in Sec. \ref{s4}. the error bars are the mean square deviations of the values obtained in different runs, varying also the number of particles in the system}
 \label{fig4}
\end{figure}
\end{center}

In Fig. \ref{fig4}  the coefficient $\mu^{*} \equiv  \mu n / \kappa_{0} T \equiv \widetilde{\mu}2 \ell n v_{0}(T)/\kappa_{0} $, where $\kappa_{0}$ is the elastic limit of the thermal heat conductivity, is plotted as a function of the restitution coefficient $\alpha$ for $d=2$ and $d=3$. The symbols are results from the simulations using Eq.\ (\ref{5.3}) to compute $\widetilde{\mu}$. For each value of $\alpha$ several values of $N_{z}$ have been considered. Besides, different realizations have been run in each case. The error bars in the figure are the standard deviation of the obtained values. The solid lines are the theoretical prediction given by Eq.\ (\ref{4.13}), which is  indistinguishable from the results reported in refs, \cite{BDKyS98} and \cite{ByC01}. Equation (\ref{5.2}) has also been used in ref. \cite{RyS03} to measure the transport coefficient $\mu$, but restricting the analysis to the quasi-elastic region. Moreover, profiles very similar to those in Fig. \ref{fig3} have been observed in experiments \cite{CyR91,WHyP01,ByK03,HYCMyW04}, although in some of them it was not associated to the modified heat current discussed here.

Finally, let stress that the purpose here was not to argue about the existence of the term proportional to the density gradient in the modified Fourier law, but to provide some insight into the physical mechanisms leading to it, and to illustrate some observable consequences of the new term. It is worth to emphasize the crucial role played by the inelasticity of collisions and the associated energy dissipation in the modification of the Fourier law. It is the time dependence of the reference homogeneous distribution function rather than the presence of large temperature or density gradients that is the origin of the new term in the expression of the heat flux.

\begin{theacknowledgments}

Comments from M.I. Garc\'{\i}a de Soria and P. Maynar are gratefully acknowledged. This research was supported by the Ministerio de Econom\'{\i}a y Competitividad (Spain)
through Grant No. FIS2011-24460 (partially financed by FEDER funds).

\end{theacknowledgments}

\bibliographystyle{aipproc}

\end{document}